\documentclass[reprint,aps,amsmath,amssymb,nofootinbib,twoside,prd,showkeys,superscriptaddress]{revtex4-1}
\pdfoutput=1
\usepackage{verbatim}
\usepackage{comment}
\usepackage{graphicx}
\usepackage[T1]{fontenc}
\usepackage{epsfig}
\usepackage{bm}
\usepackage{tensor}
\usepackage{amssymb}
\usepackage{float}
\usepackage{amsmath}
\usepackage{subfigure}
\usepackage{dcolumn}
\usepackage[utf8]{inputenc}
\usepackage{cancel}
\usepackage[colorlinks]{hyperref}
\usepackage[usenames,dvipsnames]{color}
\hypersetup{
     breaklinks=true,
    pdfstartview={FitH},    
    colorlinks=true,       
    linkcolor=blue,          
    citecolor=red,        
    filecolor=magenta,      
    urlcolor=blue,           
    anchorcolor=green,      
    linktocpage=true
}

\def\doi{http://doi.org}

\newcommand{\abs}[1]{\vert{#1}\vert}

\newcommand{\HCd}{\mathcal{H}}

\def\HCdt0{\tilde{\HCd}_{0}}

\newcommand{\affcam}{DAMTP, Centre for Mathematical Sciences, University of Cambridge, Wilberforce Road, Cambridge CB3 0WA, United Kingdom}
\newcommand{\affcamast}{Kavli Institute of Cosmology (KICC), University of Cambridge, Madingley Road, Cambridge, CB3 0HA, UK}

\newcommand{\affbulg}{Institute for Nuclear Research and Nuclear Energy, Bulgarian Academy of Sciences, Sofia, Bulgaria}
\newcommand{\affmaltaa}{Institute of Space Sciences and Astronomy, University of Malta, Malta, MSD 2080}
\newcommand{\affmaltab}{Department of Physics, University of Malta, Malta, MSD 2080}

\begin{document}

\title{On the Robustness of the Constancy of the Supernova Absolute Magnitude: Non-parametric Reconstruction \& Bayesian approaches }

\author{David Benisty}
\email{db888@cam.ac.uk}
\affiliation{\affcam}\affiliation{\affcamast}
\author{Jurgen Mifsud}
\email{jurgen.mifsud@um.edu.mt}
\affiliation{\affmaltaa}\affiliation{\affmaltab}
\author{Jackson Levi Said}
\email{jackson.said@um.edu.mt}
\affiliation{\affmaltaa}\affiliation{\affmaltab}
\author{Denitsa Staicova} 
\email{dstaicova@inrne.bas.bg}
\affiliation{\affbulg}

\begin{abstract}
In this work, we test the robustness of the constancy of the Supernova absolute magnitude $M_B$ using Non-parametric Reconstruction Techniques (NRT). We isolate the luminosity distance parameter $d_L(z)$ from the Baryon Acoustic Oscillations (BAO) data set and cancel the expansion part from the observed distance modulus $\mu(z)$. Consequently, the degeneracy between the absolute magnitude and the Hubble constant $H_0$, is replaced by a degeneracy between $M_B$ and the sound horizon at drag epoch $r_d$. When imposing the $r_d$ value, this yields the $M_B(z) = M_B + \delta M_B(z)$ value from NRT. We perform the respective reconstructions using the model independent Artificial Neural Network (ANN) technique and Gaussian processes (GP) regression. For the ANN we infer $M_B = -19.22\pm0.20$, and for the GP we get $M_B = -19.25\pm0.39$ as a mean for the full distribution when using the sound horizon from late time measurements. These estimations provide a $1\,\sigma$ possibility of a nuisance parameter presence $\delta M_B(z)$ at higher redshifts. We also tested different known nuisance models with the Markov Chain Monte Carlo (MCMC) technique which showed a strong preference for the constant model, but it was not possible not single out a best fit nuisance model.
\end{abstract}

\maketitle

\section{Introduction}
The $\Lambda$CDM model is a six parameter model that is not only the simplest parametric model of cosmology but also the one that has historically been the most successful in explaining cosmological observations. In the most recent reporting, $\Lambda$CDM has shown to be in an excellent agreement with measurements of the cosmic microwave background (CMB) radiation, the abundances of elements in the early Universe, the large scale structure of the Universe and other major astronomical measurables \cite{Planck:2018vyg,eBOSS:2020yzd,Heymans:2020gsg,Efstathiou:2019mdh}. While foundational problems persist in $\Lambda$CDM such as the nature of dark matter and dark energy, as well as long-standing theoretical issues in general relativity (GR) \cite{Weinberg:1988cp}, the fortitude of the concordance model has prompted new questions in the foundations of basic cosmological principles \cite{Krishnan:2021dyb}.

The revisiting of the foundations of $\Lambda$CDM has come about due to the growing observational crisis primarily centered on the value of the Hubble constant $H_0$, which appears to be the most serious of the emerging cosmological tensions \cite{DiValentino:2020vhf,DiValentino:2020zio,DiValentino:2020vvd,DiValentino:2020srs}. The discrepancy between early and late time measurements of the value of $H_0$ came to the fore with the reporting of local values of $H_0$ from the SH0ES \cite{Riess:2021jrx,Riess:2019cxk}, the latest of which gives $H_0^\mathrm{S21} = 73.04 \pm 1.04\,{\rm km}\, {\rm s}^{-1}\, {\rm Mpc}^{-1}$ \cite{Riess:2021jrx}, which are brought down by strong lensing measurements \cite{Wong:2019kwg} 
and measurements from the tip of the red giant branch \cite{Freedman:2019jwv,Freedman:2021ahq}. On the other end of the spectrum, early Universe measurements give a drastically lower value with the latest Planck Collaboration value being $H_0^\mathrm{Pl18} = 67.4 \pm 0.5\,{\rm km}\, {\rm s}^{-1}\, {\rm Mpc}^{-1}$ \cite{Planck:2018vyg}. Other early time measurements are largely consistent with this value such as the extended Baryon Oscillation Spectroscopic Survey (eBOSS) \cite{eBOSS:2020yzd}, the Dark Energy Survey \cite{DES:2017txv}, and results from the Atacama Cosmology Telescope (ACT) \cite{ACT:2020gnv}. The $5\sigma$ disagreement between the early time estimation and the late time estimation of $H_0$ is so dramatic that it represents one of the biggest challenges in modern cosmology. 

Another approach to viewing this tension is through the sound horizon at last scattering, $r_d$, which depends on the matter content of the pre-recombination Universe and which can have an enormous impact on the way we interpret expansion data \cite{Knox:2019rjx}. In this work, we will utilize both the Planck Collaboration fiducial value of the sound horizon, $r_d^\mathrm{Pl18} = 147.09\pm0.26\,\mathrm{Mpc}$ \cite{Planck:2018vyg,Gomez-Valent:2021hda}, and the late time value given by $r_d^\mathrm{HW+SN+BAO+SH0ES} = 136.1\pm2.7\,\mathrm{Mpc}$ \cite{Arendse:2019hev}.

The reaction to the cosmological tension has been varied with numerous proposals for solutions \cite{DiValentino:2021izs,Beenakker:2021vff} with a large portion of the community still seeking some systematics source to the cosmological tensions problem. However, these efforts are diminished with every new data release. By and large, the efforts to resolve the tension fall into the following possible modifications: departures from GR on cosmic scales \cite{CANTATA:2021ktz,Clifton:2011jh,Addazi:2021xuf,Bahamonde:2021gfp,Vagnozzi:2019ezj,Takahashi:2021bti}; adding new species to the $\Lambda$CDM model such as those described in Refs. \cite{DiValentino:2021imh,DiValentino:2021izs}; radical new physics at recombination that alters the sound horizon~\cite{Karwal:2016vyq, Poulin:2018cxd,Sakstein:2019fmf,Smith:2019ihp}; or alterations to the expansion history at late times~\cite{DiValentino:2019ffd,DiValentino:2019jae,Martinelli:2019krf,Menote:2021jaq}.

In this work, we test the robustness of this statement against type Ia supernovae (SNIa) and baryonic acoustic oscillation (BAO) data. This may provide another avenue by which the tension in the Hubble diagram may be interpreted. Our approach involves taking several parametric models of $M_B = M_B(z)$, together with two non-parametric approaches. In the parametric case, we take a number of literature models for the possible variation in absolute magnitude \cite{Ferramacho:2008ap,Linden:2009vh,Tutusaus:2017ibk,DiValentino:2020evt,Mazo:2022auo}. On the other hand, for the non-parametric approaches we test for possible dependence on redshift through two independent machine learning probes, namely Gaussian processes (GPs) and artificial neural networks (ANNs) which we expand upon later on. 

The plan of the work is as follows: Section \ref{sec:DistLad} reconstructs $M_B$ from these measurements, while in Section \ref{sec:GP} and \ref{sec:ANN} we adopt the GPs and the ANNs for the reconstruction of $M_B$, respectively. In Section \ref{sec:ModCom} we use the Bayesian analysis technique for different $\delta M_B$ models, while we discuss our reported results in section \ref{sec:Dis}.

\begin{table} 
\scalebox{1.1}{
\begin{tabular}{|c|c|c|c|c|}       
\hline\hline                                                                                                    

$z$  & $D_A/r_d$ & $\sigma_{Data}$ & year  &  Ref. \\ \hline\hline 
$0.11$   & $2.607$ & $0.138$&  $2021$  & \cite{deCarvalho:2021azj}\\
$0.24$    & $5.594$& $0.305$&  $2016$  & \cite{BOSS:2016goe}\\
$0.32$     &  $6.636$  &  $0.11$   &  $2016$   & \cite{BOSS:2016wmc} \\
$0.38$     &  $7.389$  &  $0.122$  &  $2019$  &  \cite{BOSS:2016hvq}\\
$0.44$     &  $8.19$  &  $0.77 $  &   $2012$   & \cite{Blake:2012pj}\\
$0.51$    &  $7.893$  &  $0.279$  &   $2015$ & \cite{Carvalho:2015ica}	\\
$0.54$     &  $9.212$  &  $0.41$    &  $2012$  & \cite{ Seo:2012xy}\\
$0.6$     &  $9.37$  &  $0.65$   &  $2012$  & \cite{Blake:2012pj}\\
$0.697$     &  $10.18$  &  $0.52$   &  $2020$  & \cite{Sridhar:2020czy}\\
$0.73$     &  $10.42$  &  $0.73$  &   $2012$   & \cite{Blake:2012pj}\\
$0.81$     &  $10.75$  &  $0.43$   &  $2017$   & \cite{DES:2017rfo}\\
$0.85$     &  $10.76$  &  $  0.54$   &  $ 2020$  & \cite{Tamone:2020qrl}\\
$0.874$     &  $11.41$  &  $0.74$    &  $2020$ & \cite{Sridhar:2020czy}\\
$1.00$     &  $11.521$  &  $1.032$    &  $2019$ & \cite{Zhu:2018edv}\\
$1.480$     &  $12.18$  &  $0.32$  &  $2020$  & \cite{Hou:2020rse}\\
$2.00$     &  $12.011$  &  $0.562$  &   $2019$  & \cite{Zhu:2018edv}\\
$2.35$     &  $10.83$  &  $0.54$  &  $2019$   & \cite{Blomqvist:2019rah}  \\
$2.4$     &  $10.5$  &  $0.34$  & 2017 & \cite{duMasdesBourboux:2017mrl}\\

\hline\hline                                                
\end{tabular} 
}
\caption{ {\it{A compilation of angular BAO measurements from luminous red and blue galaxies and quasars from diverse releases of the SDSS.}}}
\label{tab:data}                                            
\end{table}  

\section{Distance Ladders}
\label{sec:DistLad}
A full resolution to the tension remains largely an open problem. In this work, we attempt to probe the nuanced features of the Hubble tension using late time expansion data by exploring the constancy of the absolute magnitude of  SNIa. We start with the luminosity distance which is fixed by the evolution of the Hubble parameter through
\begin{equation}
	d_L(z) = (1+z)  \int_0^z \frac{c\,dz'}{H(z')}\,,
\end{equation}
where $c$ is the speed of light, and which naturally leads to the angular diameter distance $D_A(z)=d_L(z)/(1+z)^2$. Thus, irrespective of the source of the corrections needed to resolve the Hubble tension, these measurements can constrain the evolution of $H(z)$. We also note that it is not possible to include CMB data without assuming a cosmological model and so we only consider late time data to avoid adding this further constraint on our hypothesis. We can thus retain independence from a cosmological model throughout the GPs and ANN reconstructions.

For the SNIa standard candles, the distance modulus $\mu_{Ia}^{}(z)$ (defined as the difference between absolute and relative magnitudes) is related to the luminosity distance through
\begin{equation}
	\mu_{Ia}^{} (z) = 5 \log_{10} \left[ d_L(z)\right] + 25 + M_B (z) \,,\label{eq:dist_mod_def}
\end{equation}
where $d_L(z)$ is measured in units of Mpc, and where $M_B(z)$ represents the possible redshift dependence of the intrinsic magnitude, which is regularly assumed to be a constant. In the present work, we explore the possibility that this may inherit some dynamics due to the cosmic evolution across redshift. We do not speculate on the potential sources, whether astrophysical or cosmological, for such a redshift dependence to arise in this work. The emergence of any possible redshift dependence would be sourced from the absolute magnitude, while $M_B(z)$ is largely assumed to be a constant in most of the literature. 

The possibility of a redshift variation in the absolute magnitude was suggested in Refs.~
\cite{Camarena:2019moy, Camarena:2021jlr} 
where they claim that the Hubble tension can also be reinterpreted through these variations in $M_B(z)$. Indeed, Refs.~\cite{Marra:2021fvf,Alestas:2020zol,Alestas:2021xes,Alestas:2021luu,Dainotti:2021pqg,2110.05346} suggest that a transition in the dark energy equation of state or in the absolute magnitude as a possible solution to the Hubble tension. In addition to the SNIa observations, we also make use of BAO measurements \cite{Blake:2012pj,Blake:2012pj,Seo:2012xy,Blake:2012pj,Carvalho:2015ica,BOSS:2016goe,BOSS:2016wmc,BOSS:2016hvq,deCarvalho:2017xye,duMasdesBourboux:2017mrl,Zhu:2018edv,DES:2017rfo,Blomqvist:2019rah,Sridhar:2020czy,Sridhar:2020czy,Tamone:2020qrl,Hou:2020rse,deCarvalho:2021azj, Camarena:2019rmj} in order to take a fuller account of the possible evolution of $M_B(z)$. We incorporate BAO measurements into Eq.~\eqref{eq:dist_mod_def} by substituting in for the angular diameter distance giving
\begin{subequations}
\begin{equation}
    M_B = \mu_{Ia}^{} - 5\log_{10}\left[ (1+z)^2 \left(\frac{D_A^{}}{r_d^{}}\right)_\mathrm{BAO}\cdot r_d^{} \right] - 25\,,
\label{eq3a}
\end{equation}
where the reciprocal theorem was used, and which gives uncertainties
\begin{equation}
    \Delta M_B^{}  = \Delta\mu_{Ia}^{} + \frac{5}{\ln 10}\left[\frac{\Delta r_d^{}}{r_d^{}} + \frac{\Delta \left(D_A^{}/r_d^{}\right)_{BAO}}{\left(D_A^{}/r_d^{}\right)_{BAO}}\right]\,.
\end{equation}
\end{subequations}

Since Eq. (3b) is an analytical derivation of the error for Eq. (3a), it does not make an assumption on the Gaussianiaty of the data. Furthermore, here $r_d$ enters as a measurement with its mean and error and not as a prior.

We reconstruct the $\mu_{Ia}^{}$ from the Type Ia supernova and the $\left(D_A/r_d\right)_\mathrm{BAO}$ from the BAO measurements (see Table \ref{tab:data}). Through these reconstructions, we are able to produce intermediary points for both types of data, so that be used to assess to what level the $M_B$ nuisance parameter adheres to being a constant. As a consequence of this approach, the degeneracy between the $M_B$ and $H_0$, is replaced by $M_B$ and $r_d$. The BAO data set we use does not include covariance matrices but instead tests for the effect of possible correlations on the final results (for more details see Ref. \cite{Benisty:2021gde}).

\begin{figure*}[t!]
 	\centering
\includegraphics[width=0.7\textwidth]{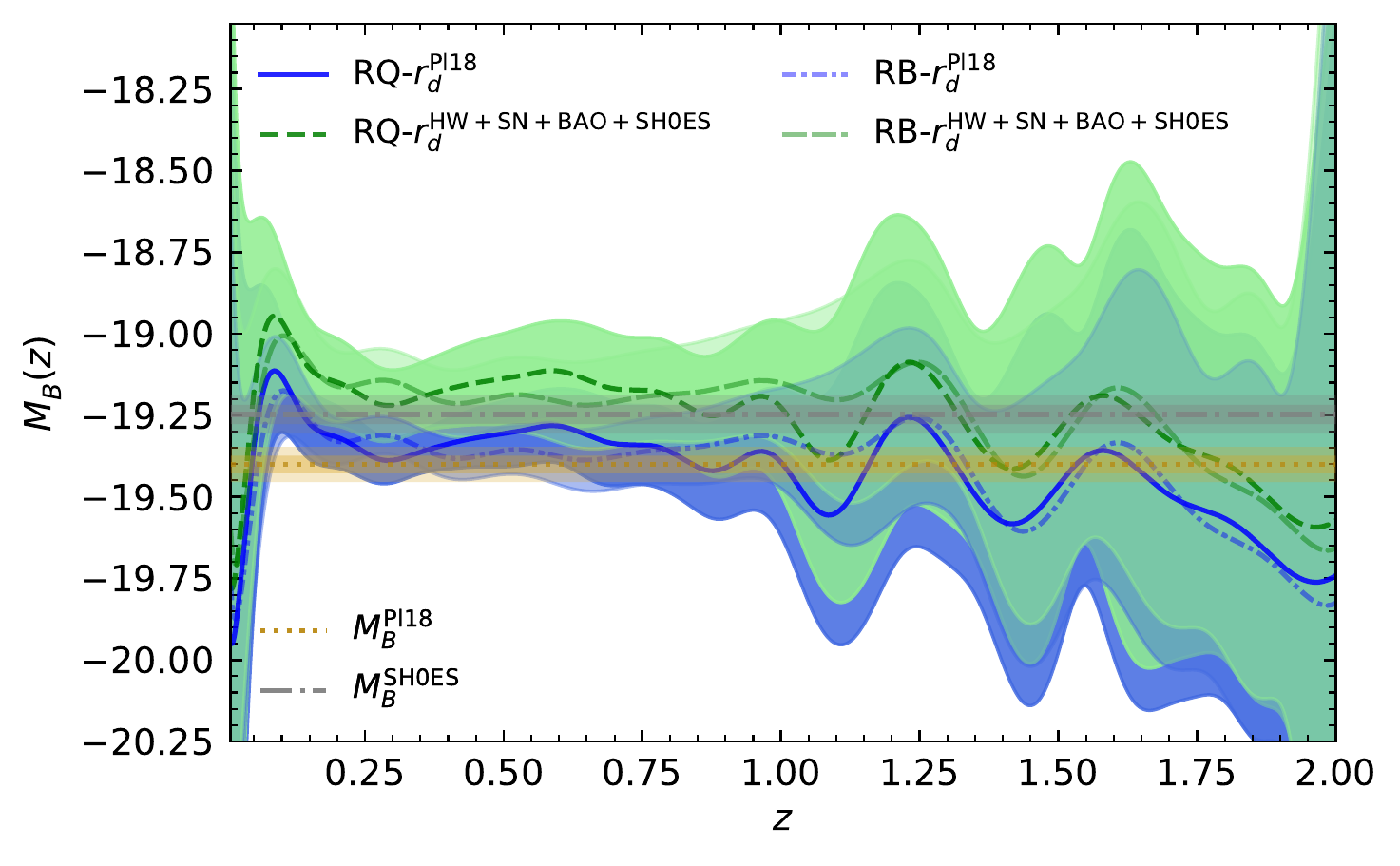}\\
\includegraphics[width=0.7\textwidth]{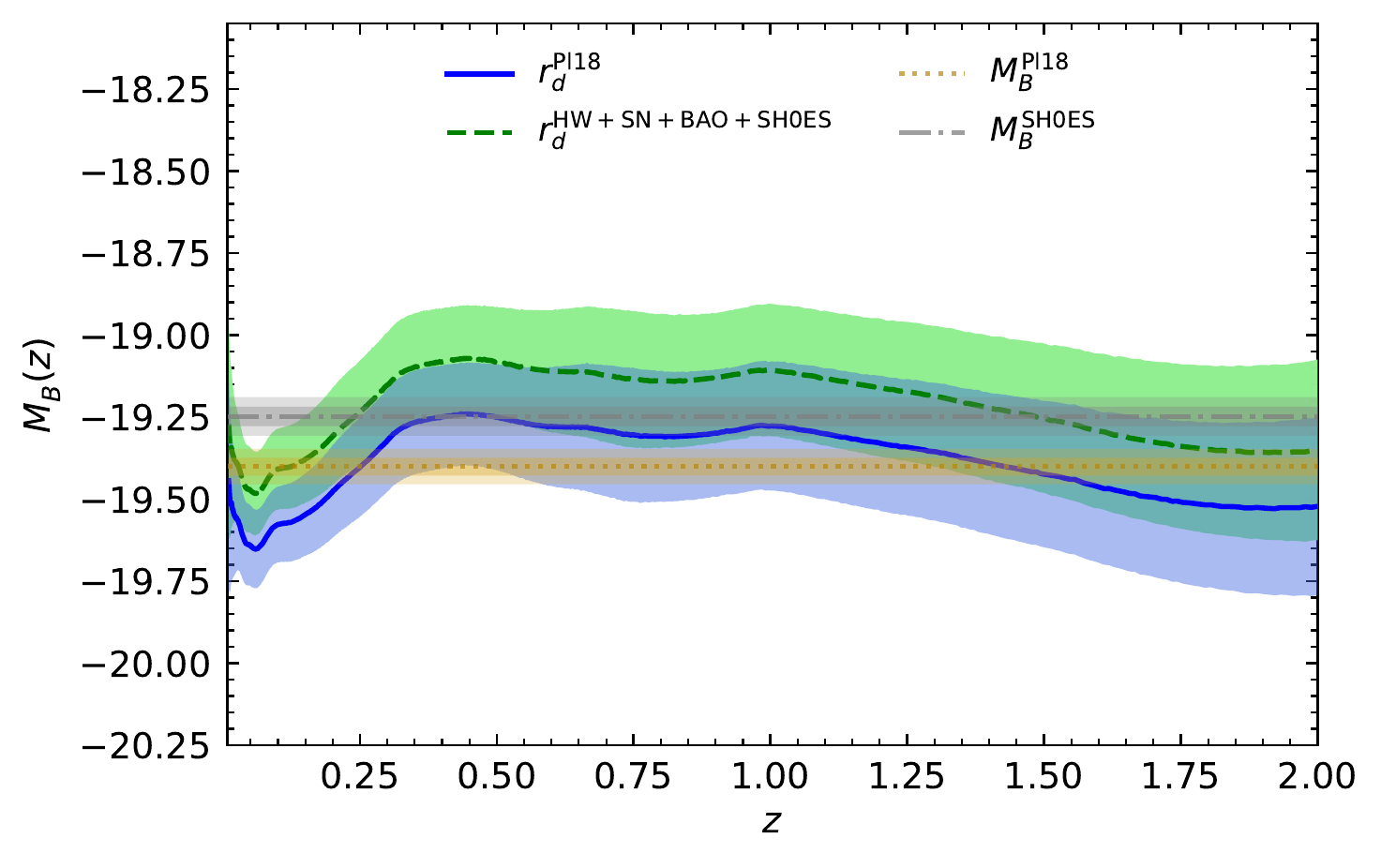}
\caption{\it{Nonparametric Reconstruction for the Absolute Magnitude from GPs (upper) and ANN (lower) with a 68\% confidence interval. The kernels are Rational Quadratic (RQ) and Radial Basis (RB) function. The reconstructions of $M_B(z)$ when assuming the sound horizon $r_d$ from the Planck (blue) and the H0LiCOW+SN+BAO+SH0ES (green) are illustrated in both panels. }}\label{fig:ReconM}
\end{figure*}

\section{Gaussian Process Reconstruction} \label{sec:GP}

The GP reconstructs the data set elements as part of a stochastic process in which each element is the taken to be part of a multivariant normal distribution \cite{10.5555/971143,10.5555/1162254,Renzi:2020fnx}. GPs have been applied in a number of cosmological scenarios \cite{Shafieloo:2012ht,Seikel:2013fda,Yang:2015tzc,Cai:2015pia,Wang:2017jdm,Zhou:2019gda,Mukherjee:2020vkx,Gomez-Valent:2018hwc,Zhang:2018gjb,Aljaf:2020eqh,Li:2019nux,Liao:2019qoc,Yu:2017iju,Yennapureddy:2017vvb,Benisty:2020kdt,Briffa:2020qli,LeviSaid:2021yat,Escamilla-Rivera:2021rbe,Bernardo:2021cxi,Bernardo:2021qhu,Bernardo:2021mfs,Cai:2019bdh,Colgain:2021ngq,ruizzapatero2022modelindependent,Ren:2021tfi,Singirikonda:2020ieg}. The GPs will be defined via a mean function $\mu(z)$, and a kernel, function $k(z,\tilde{z})$ which together describe the continuous realisation of the GPs reconstruction $\xi(z) \sim \mathcal{GP} \left(\mu(z), k(z,\tilde{z})\right)$, where $\tilde{z}$ represents the elements of the input data sets. 

GPs utilizes a Bayesian approach to optimizing its kernel hyperparameters leaving open the functional choice of the kernel, so that they are fit using Bayesian optimizer approach where the difference between the reconstructed and observed behaviors is minimized at the redshift points where observations exists. For any two redshift points $z$ and $\tilde{z}$, the kernel incorporates the strength of the correlation for the reconstructed parameter \cite{Busti:2014aoa}. There exist a number of literature kernel choices \cite{10.5555/1162254,Seikel:2013fda} which were designed to be general purpose, but which generally agree with each other to within an amount of uncertainty. In this work we consider the Radial Basis (RB) function kernel:
\begin{equation}\label{eq:cov-squ}
    k(z,\tilde{z}) = \sigma_f^2\exp\left(-\frac{(z-\tilde{z})^2}{2l^2}\right)\,,
\end{equation}
and the Rational Quadratic (RQ) kernel:
\begin{equation}\label{eq:}
    k(z,\tilde{z}) = \frac{\sigma_f^2}{\left(1 + \abs{z-z'}^2/2\alpha l^2 \right)^\alpha}\,.
\end{equation}
We have also tested the Matern kernel, but we exclude these results from the figures since they give identical results to those obtained with the RQ kernel.

These kernels are infinitely differentiable and represent different expressions of the hyperparameters $\sigma_f$ and $l$ which characterize the smoothness and overall profile of the reconstructed profile. The length-scale $l$ signifies the distance to which pairs of elements in the data set can influence each other to a significant enough extent, while $\sigma_f$ controls the uncertainties across the redshift range of the reconstruction. 

The upper panel of Fig. \ref{fig:ReconM} shows the final reconstruction of $M_B\left(z\right)$ for both choices of $r_d$, namely the one measured by Planck 2018 and the one inferred from the H0LiCOW+SNIa+BAO+SH0ES measurement. Both the RQ and the RB kernels show the same behavior of $\delta M_B$ at higher redshifts.

For the low redshift, the GPs predicts a singularity for $z=0$ which comes from a numerical artifact, since we take the quantity $\sim \log_{10} \left(D_A(z)\right)$ and for $z = 0$ the $D_A$ is zero. The fluctuations for $z>1$ may be caused by the the BAO data set possessing only a small number of measurements in this redshift range. However, the possibility for $\delta M_B(z) < 0$ remains.

\begin{figure*}[t!]
 	\centering
    \includegraphics[width=0.85\textwidth]{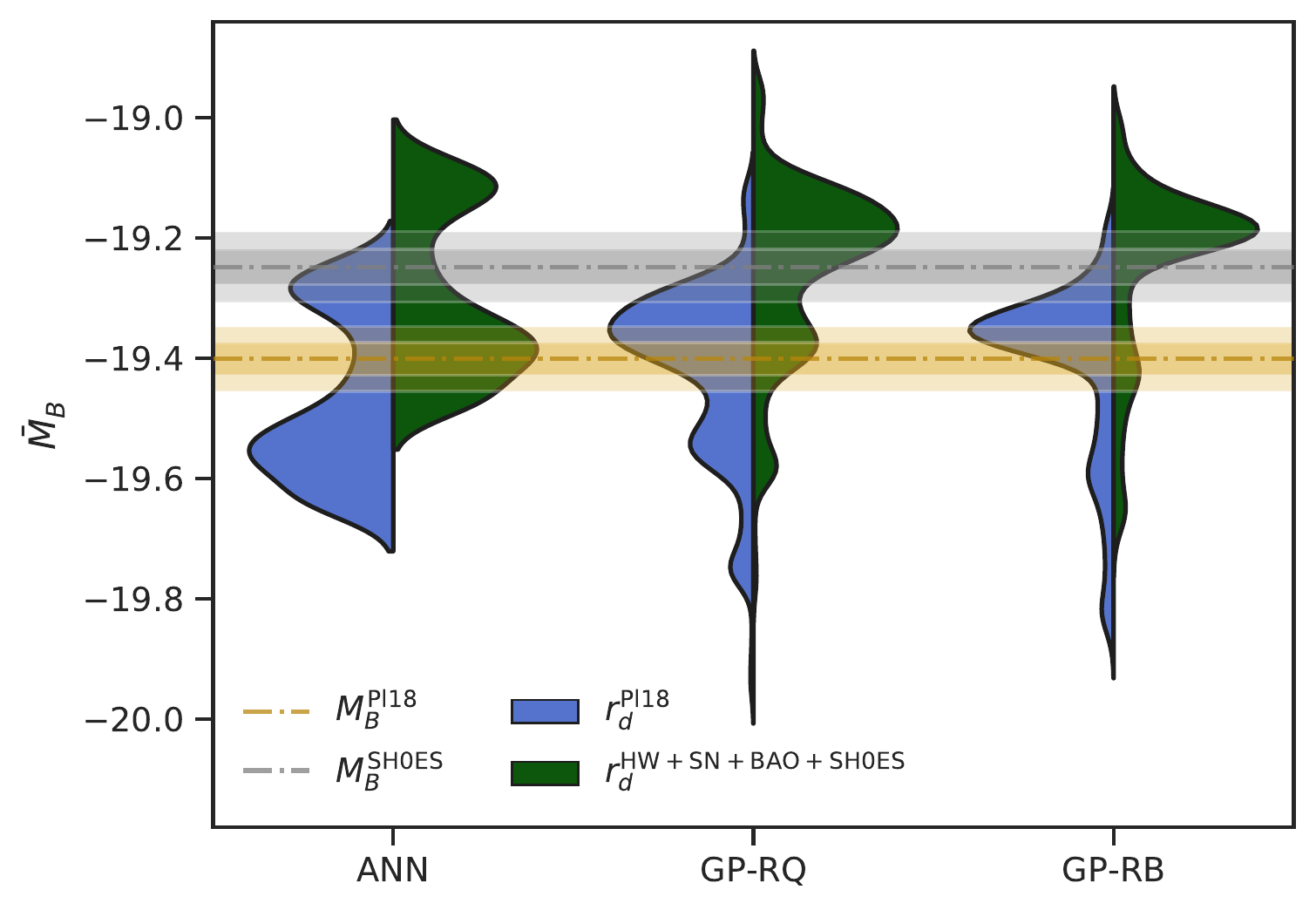}
    \caption{\it{Distribution of the mean ANN and GPs reconstruction values of the absolute magnitude $\bar{M}_B^{}$, with the considered fiducial values of $r_d^{}$. The blue distribution presents the calibrated $\bar{M}$ with the Planck estimated sound horizon and the green distribution depicts the calibrated $\bar{M}$ with the sound horizon from the late time measurements (H0LiCOW+SN+BAO+SH0ES).}}\label{fig:violin}
\end{figure*}

The new degeneracy $M_B - r_d$ from the combination Type Ia + BAO can yield the value of $M_B(z)$ assuming a particular $r_d^{}$. The mean of $M_B(z)$ (over the two kernels) gives $M_B = -19.42\pm0.35$ for the sound horizon from $r^{\rm Pl18}_d$, and $M_B = -19.25\pm0.39$ for the sound horizon using H0LiCOW+SN+BAO+SH0ES. Different kernels give only a different error, see Table \ref{tab:r_d_vals}. The results are consistent with latest measurement of $M_B$ for the sound horizon by H0LiCOW+SN+BAO+SH0ES. The projected $M_B$ values are characterised by a small tail distribution with one peak, as  illustrated in Fig.~\ref{fig:violin}.

\section{Artificial Neural Network Reconstruction} \label{sec:ANN}

ANNs offer a promising method of modeling natural processes \cite{2015arXiv151107289C,Dialektopoulos:2021wde}. The ANN approach is comprised of input and output layers which connects the internal, or hidden, layers to the respective input and output data. The hidden layers give structure to the ANN neurons whose relationships define how the ANN will respond to different inputs. In our scenario we are interested in input redshifts and output Hubble parameters together with their associated uncertainties. The process by which an ANN gives an output is a convoluted one with each neuron accepting any number of inputs but only outputting a single value. The output of a neuron is defined by a weighted sum of its inputs which may be complemented by a bias term, this is then passed through a nonlinear transformation (or activation function) to produce the neuron's output. The initial input and output layers perform linear transformations on the incoming and outgoing signals. In this way, an input signal will traverse the entire network in a structured manner.

The ANN features a huge number of other hyperparameters, such as the weighted sum, which are optimized during the learning process to better model the training data. The way in which the number of neurons in an ANN are selected and structured into layers needs to be optimized using a subset of the observational data or, as in our case, the use of mock data sets which are generated by mimicking the observational redshift distribution and with an identical number of observational data points. We do this using a risk-like statistic in which for each trained ANN design, we compare the profile for increasing number of neurons for every number of hidden layers (for example, one to three layers). For the mock data, for instance in the case of the SNIa $\bar{\mu}_{Ia}^{}(z_i)$, we compare the ANN output predicted values $\mu_{Ia}^{}(z_i)$ together with their associated ANN uncertainties $\sigma_{Ia}^{}(z_i)$ through \cite{Wasserman:2001ng}
\begin{align}
    \mathrm{risk} &= \sum_{i=1}^N\mathrm{bias}_i^2+\sum_{i=1}^N\mathrm{variance}_i^{} \nonumber\\
    &=\sum_{i=1}^N\left[\mu_{Ia}^{}(z_i)-\bar{\mu}_{Ia}^{}(z_i)\right]^2+\sum_{i=1}^N\sigma^2_{Ia}\left(\mu_{Ia}^{}(z_i)\right)\,,
\end{align}
where $N$ is the total number of SNIa input data points. This gives a clear method by which we select the optimal number of layers and neurons.

The other seemingly ambiguous part of this reconstruction technique is the approach by which the ANN is trained with real data. The number of iterations through which the ANN is successively trained is determined by a loss function which quantifies directly the difference between predicted values and the ground truth for the training data. Similar to the GP optimizer, ANNs are optimized against observational data using the loss function since it quantifies the differences between the reconstructions and the observed data at redshift points where observations are present. There are a variety of possible loss functions (see, for instance, Ref.~\cite{2015arXiv151107289C}) such as the L1 loss function. The L1 loss function is robust against outliers in the data but is not very stable for small adjustments in the data which possibly gives multiple solutions. For this reason we also consider the L2 loss function which takes a mean square error approach to quantifying the fit of the ANN predicted values. While the L2 loss function is less robust to outliers, it is more stable for small changes to ANN predictions and gives unique solutions. In each iteration, the hyperparameters of the ANN are updated through a gradient--based optimizer. In this work, we take Adam's algorithm \cite{2014arXiv1412.6980K} as our optimizer since this will accelerate the training convergence. In our analyses, the L1 loss function was found to be the optimal loss function for both the Pantheon and BAO data sets. Moreover, in the case of the Pantheon data set we adopted an ANN with three layers and 256 neurons, whereas for the BAO data set an ANN consisting of two layers and 8192 neurons was found to be the optimal network structure.

\begin{table*}[t!]
\centering
\begin{tabular}{|c|c|c|c|c|}
\hline \hline
        Technique & $r_{d,fit}^\mathrm{Pl18}$ & $r_{d, fit}^\mathrm{HW+SN+BAO+SH0ES}$ & $r_{d,full}^\mathrm{Pl18}$ & $r_{d,full}^\mathrm{HW+SN+BAO+SH0ES}$ \\
        \hline\hline 
        
  ANN  & $-19.58 \pm 0.11$ and $-19.26 \pm 0.04$ &  $-19.1 \pm 0.04$ and $-19.42 \pm 0.11$& $-19.38\pm 0.20$ &$-19.22\pm 0.20$ 
  \\
        \hline
 GP-RQ & $-19.35 \pm 0.03$  & $-19.18 \pm 0.03$ & $-19.42 \pm0.35$ & $-19.25 \pm 0.39$\\
        \hline
 GP-RB &  $ -19.35\pm 0.07$ & $-19.18 \pm 0.07$ & $-19.42\pm 0.29$  & $-19.25\pm 0.33$ \\
        \hline  \hline
\end{tabular}
\caption{\label{tab:r_d_vals}\it{The inferred ANN and GPs mean values of $M_B^{}(z)$ are shown here when adopting $r_d^\mathrm{Pl18}$ and $r_d^\mathrm{H0LiCOW+SN+BAO+SH0ES}$ fiducial values. The values for the ANN are fitted for a double Gaussian and the values for GP are fitted for a Gaussian without a tail. The final two columns give the mean and error for the full distribution }}
\label{ANNGP}
\end{table*}

ANNs with at least one hidden layer and a continuous non--linear activation function can approximate any continuous function \cite{HORNIK1990551} making this an appropriate approach to cosmological data. In this work, we use the open-source \texttt{PyTorch}\footnote{\url{https://pytorch.org/docs/master/index.html}} based code for reconstructing functions from data called Reconstruct Functions with ANN (\texttt{ReFANN}\footnote{\url{https://github.com/Guo-Jian-Wang/refann}}) \cite{Wang:2019vxv} which was implemented on GPUs. To accelerate this process, we also make use of batch normalisation \cite{2015arXiv150203167I} which is implemented prior to every nonlinear layer to effects significantly the convergence time and stability of the ANN.

The results from the ANN reconstruction can be seen in Fig.~\ref{fig:ReconM}. Here, the contrast with the GP reconstructions can be clearly observed. As in the GP case, we plot the reconstructions that result from considering Eq.~\eqref{eq3a} for two prior values of $r_d$. In this case, we consider one format of the ANN while in the GP case we consider two kernels to show consistency between these choices. One can clearly observe that the ANN eliminates the drastic fluctuations from the reconstructed function, i.e. in a way it clears the "noise", as compared with the GPs reconstruction. However, it leaves the over-all trend of the "jump" in small $z$ ($z\in (0.25, 0.5)$) and also the notable decrease in the value of $M_B(z)$ for high $z$. In addition, there is an apparent difference in the means of the GPs and ANN reconstructions in the neighbourhood of zero redshift. However, when taken in the context of the 1$\sigma$ uncertainties, this disappears since the GPs uncertainties here are very large due to a numerical divergence close to the origin. The Gaussians of $M_B$ from both methods may be seen on Fig.~\ref{fig:violin}. One can see that while both methods agree on the mean values for the respective $r_d$ for the full distribution, for the ANN, we observe bimodal distributions, while for the GP, we see a Gaussian with a tail. In table \ref{ANNGP}, we have presented both the mean and standard deviation for the full distributions and also when the distribution is fitted wth a bimodal distribution (for ANN) and for a Gaussian with a tail (for GP). Interestingly, the ANN seems to suggest two populations of objects, which in the GP seems looks like a tail, i.e. a single or principle peak together with a tail distribution that diminishes with the GPs reconstructions. This difference can be due to higher errors in the GP reconstruction.

\section{Model Comparison}
\label{sec:ModCom}
The question on the constancy of $M_B$ has two aspects. One of them is in connection with the Etherington's reciprocity theorem or the so called distance duality relation (DDR) (explicitly, $d_L(z)=(1+z)^2 d_A(z)$) which theoretically should hold for any metric theories of gravity where the photon number is conserved and photons travel along null geodesics. The validity of DDR has been studied and confirmed using different astronomical sources in the pioneer works by \cite{Uzan:2004my, DeBernardis:2006ii, SupernovaCosmologyProject:2008ojh, Stern:2009ep, Luzzi:2009ae, Avgoustidis:2010ju, Ma:2016bjt}. In a number of more recent works (\cite{Renzi:2021xii, Qin:2021jqy, 2020ApJ..888...32L, He:2022phb,Ma:2016bjt,Fu:2017nmw}), different datasets have been used along with various numerical methods, to answer the question if DDR holds for all redshifts and objects and if not, whether there could be a better functional form for it. The over all conclusion is that this relation holds even though there are some evidences of mild deviations, especially for models with spatial curvature.

The other aspect and the one that we focus in this section is - if it is possible that the deviation is in $M_B$ itself, namely of the possible redshift dependence of the intrinsic luminosity of SNIa. This question has also been studied in different forms in the literature  (\cite{Ferramacho:2008ap,Linden:2009vh,Tutusaus:2017ibk,DiValentino:2020evt,Perivolaropoulos:2022khd}). To check how our approach works in this case,  we will use the methods of MCMC as implemented by {\it{Polychord}} \cite{Handley:2015fda} to consider the effect of such a dependence on the cosmological parameters. We will consider the two described above datasets and minimize $\chi^2=\chi^2_{BAO}+\chi^2_{SN}$.

\begin{figure*}
 	\centering
    \includegraphics[width=0.49\textwidth]{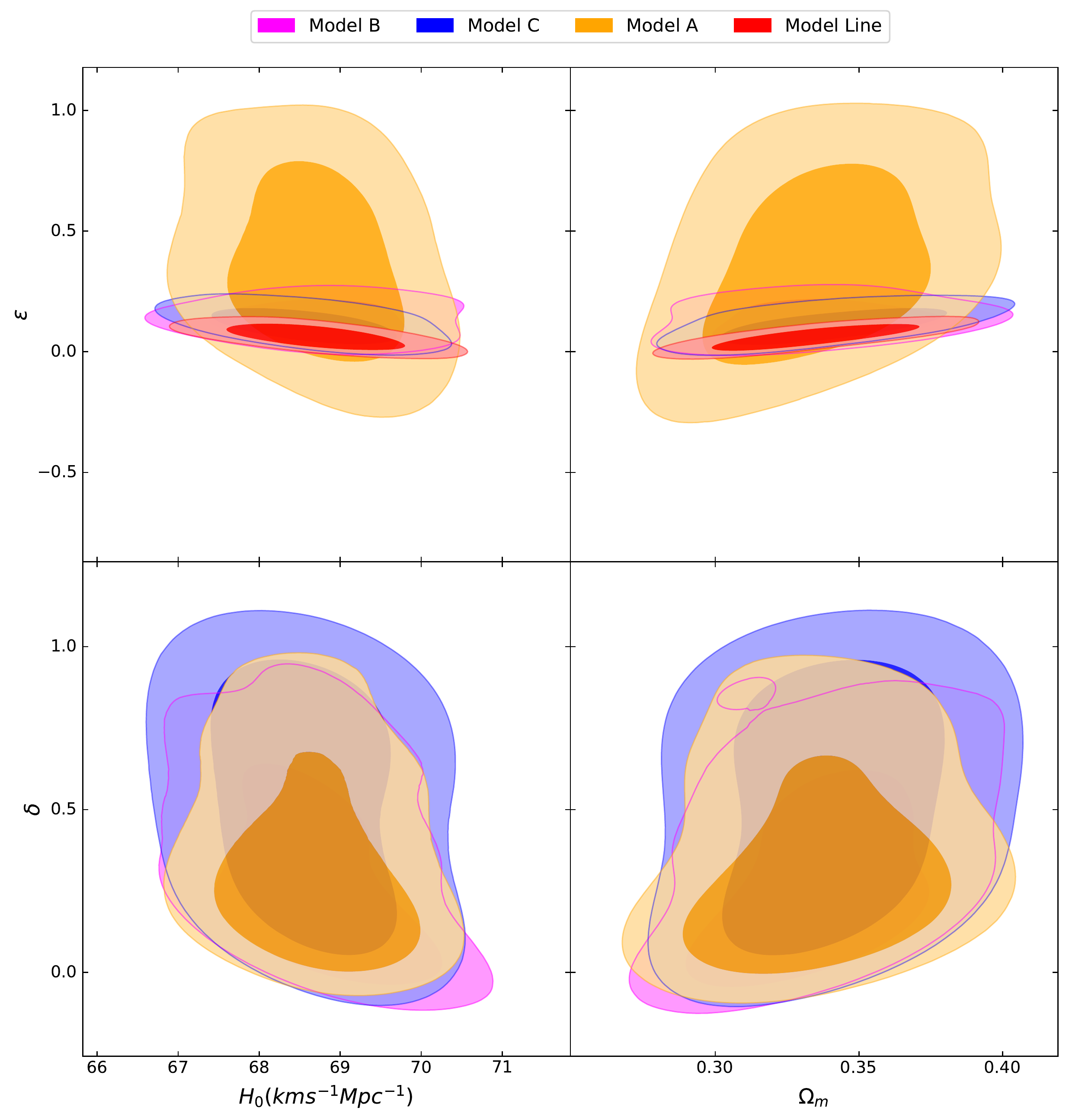}
    \includegraphics[width=0.49\textwidth]{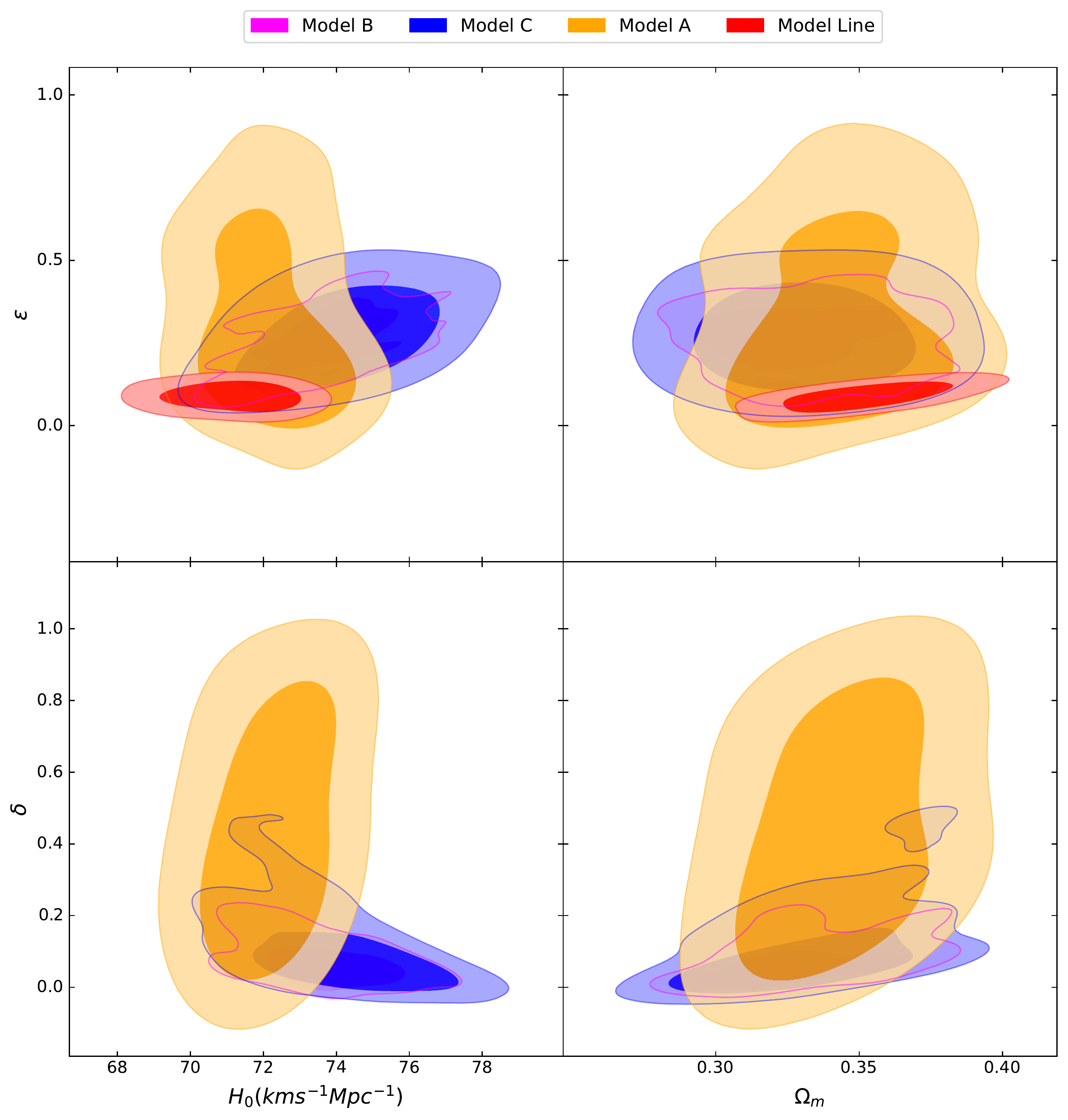}
    \caption{\it{The posterior distribution for the nuisance parameters for different models with a $\Lambda$CDM background are shown, on the right panel corresponding to a prior $r^{Pl18}_d = 147.09 \pm 0.26$ Mp, on the left one - to $r_d^{HW+SN+BAO+SH0ES} =136.1 \pm  2.7$ Mpc.}} \label{fig:BA}
\end{figure*}

In order to allow for dependency on $z$, one can modify Eq.~\eqref{eq3a} so that $M_B$ is mapped to $M_B + \delta M_B(z)$. We consider the following phenomenological models \cite{Tutusaus:2017ibk, Linden:2009vh}:
\begin{equation}
    \delta M_B(z) =
    \begin{cases}
        \epsilon z &  \text{Model Line}   \\
        \epsilon\left[(1+z)^\delta-1\right] & \text{Model A }  \\
        \epsilon z^\delta & \text{Model B }  \\
        \epsilon\left[\ln(1+z)\right]^\delta & \text{Model C }\,.
    \end{cases}
\end{equation}
We compare them with the default $M_B=const$ model. For the expansion rate we consider the flat $\Lambda$CDM model.

To account for the possible color ($C$) and time stretching of the light curve ($X_1$) corrections in the SNIa light curves, one can use the following standardization formula \cite{Tripp:1997wt}:
\begin{equation}
\mu_{obs} = m_B^* - (M_B - \alpha X_1 + \beta C),
\end{equation}  where $m_B^*$ corresponds to the observed peak magnitude in the B-band rest-frame, $M_B$ is the absolute magnitude of the SNIa in the B-band rest-frame, $\alpha$ is the the amplitude of the stretch correction and  $\beta$ is the amplitude of the color correction. The Pantheon dataset has been already pre-marginalized with respect to these two parameters and they have been taken into account into its covariance matrix \cite{ DiValentino:2020evt}. To account for the possibility that adding a redshift dependence of $M_B$ may compromise this pre-marginalization, we test our results with and without adding priors for $\alpha$ and $\beta$. We see that including the color and light curve stretching corrections as constants or as Gaussian priors, gives very similar results. We present the Gaussian ones, because they allow to account for an effect of the redshift dependence $M(z)$ on the systematics.

Another important aspect is whether to use a Gaussian prior on $H_0$ or on $M_B$. In Refs.~\cite{Efstathiou:2021ocp,Camarena:2021jlr} it was argued that it is better to use a prior on $M_B$, in order to avoid double counting of low-redshift supernovae, and to avoid the deceleration parameter being set to the standard model value $q_0 = -0.55$ (\cite{Riess:2016jrr}), while accounting for $M_B$ being constrained by local calibration. This also allows for the inclusion of more information such as the the shape of the SNIa magnitude-redshift relation, among others. The MCMC results show that the priors on $H_0$ or on $M_B$ are largely equivalent for our purposes. For this reason, we use the DES prior on $M_B$  ($M_B=-19.2478\pm 0.0294$) (\cite{Camarena:2021jlr,Marra2022}) and leave $H_0$ as a free parameter.

The other priors we use are: uniform priors on $H_0$: $\in (50,100)$, 
$\Omega_m \in (0.2,0.4)$, $\Omega_r\in (0.,0.01)$, $\epsilon^{line}\in (-1,1)$ and $\epsilon^{A,B,C}\in (-1,1)$, $\delta\in (0,1)$. We find these ranges to suitably contain the case of constant $M_B$ for each model while also offering a justifiable interval of possible variance. The priors we use for $\alpha, X_1,\beta, C$ correspond to the G10 model in Ref. \cite{Pan-STARRS1:2017jku}.

Since using a Gaussian prior on $M_B$ is equivalent to imposing Gaussian prior on $H_0$, when we choose to work with the DES prior, our remaining free parameter is $r_d$.  The two datasets we combine, the SNIa and the BAO data, are not sufficient to break the degeneracy on $r_d$, thus we explore 3 options for it: 
\begin{enumerate}
    \item One can leave $r_d$ as a free parameter with a uniform prior. In this case, the results show a tendency to get very low $H_0$ with "normal" $r_d$ or a very high $H_0$ and a very low $r_d$. This means that the system moves too freely in the $H_0-r_d$ plane. 
    
    \item One can use Gaussian priors on $r_d$ coinciding with measurements from the early and the local Universe, namely $r^{Pl18}_d = 147.09 \pm 0.26$ Mpc and $r_d^{HW+SN+BAO+SH0ES} =136.1 \pm  2.7$ Mpc, and thus to fix $H_0$. 
    
    \item One can replace the prior on $r_d$ with a prior on $\Omega_b$ by imposing a Gaussian prior on the baryonic density ($\Omega_b= 0.0224 \pm 0.0001$) and calculating $r_d$ from
\begin{equation}
    r_d = \int_{z_d}^{\infty} \frac{c_s(z)}{H(z)} dz\,,
\end{equation}
where the drag epoch $z_d \approx 1060$ is also calculated from the inferred values. The speed of sound at $z_d$ is $c_s \approx c \left(3 + 9\rho_b /(4\rho_\gamma) \right)^{-0.5}$ in the baryon-photon fluid with the baryon $\rho_b(z)$ and the photon $\rho_\gamma(z)$ densities, respectively \cite{Aubourg:2014yra}. 
\end{enumerate}

We omit presenting here case 1., because it is too uninformative. The two other cases can be seen on tables \ref{cosmo1} and \ref{selcrit}. We see that in case 2, there is still some small variation in $H_0$ and $r_d$ depending on the model, but as a whole, the inferred $H_0$ corresponds to the chosen $r_d$ and the variations in $\Omega_m$ are small. The only notable difference is in the parameters $\delta$ and $\epsilon$, which are different from zero within the 68 \% CL for all of the priors as seen from table \ref{cosmo1} in the Appendix. In case 3., $H_0$ is rather low, despite working with the DES prior on $M_B$ (the calculated $r_d> 150$).

One can use different statistical measures to compare the models. We use Akaike information criterion (AIC), Bayesian information criterion (BIC), Deviance information criterion (DIC) and the logarithm of the Bayes factor (BF) \cite{Trotta:2008qt}, where the $\Delta$ is always evaluated with respect to the zero hypothesis $M_B(z)=const$ (ex. $\Delta \mathrm{AIC}_i = \mathrm{AIC}_0 - \mathrm{AIC}_i$) and for the BF ($\mathrm{BF}=\mathrm{evidence}_0/\mathrm{evidence}_i$). The results can be seen in the table ~\ref{selcrit} in the Appendix. We find that the null hypothesis is the best model only with respect to the AIC and  BIC measurements. With respect to DIC the $M_B=const$ hypothesis is the worst, but the statistical support is weak, meaning that purely based on this measure it is not possible to distinguish between the different hypotheses. With respect to BF, we see that $M_B=const$ model is better than the Line hypothesis for both $r_d$ but the other 3 models are either indistinguishable or better than $M_B=const$. The statistical measures for $r_d^{Pl18}$ show weak support for $M_B=const$, while the measures for $r_d^{HW+SN+BAO+SH0ES}$ show support for the $M(z)$ models. In case 3), all statistical measures but DIC show some support for $M_B=const$. 

From this comparison, we can say that none of the models can be ruled out merely on statistical grounds and also none of them is strongly preferred. The reason for the weak support for the constant model may be in the large-error boxes of the data or some type of unknown systematic. But it also raises the question, is it possible that we still have to find the best functional dependence for $M_B(z)$.

\section{Discussion} 
\label{sec:Dis}
Supernovae used as distance indicators that lead to the discovery of cosmic acceleration. Cosmic acceleration is usually attributed to a form of DE. However, in recent years, several studies have raised the question of whether or not intrinsic SNIa luminosities might evolve with redshift \cite{Ferramacho:2008ap,Linden:2009vh,Tutusaus:2017ibk,Kumar:2021djt,DiValentino:2020evt}. Here we test the soundness of inferred absolute magnitude combining SNIa data with BAO measurements. We isolate the luminosity distance parameter $d_L(z)$ from the BAO dataset and cancel the expansion part from the observed distant modulus $\mu_{Ia}^{}(z)$. Consequently, the combination with the BAO replaces the known degeneracies between $M_B$ and $H_0$ for the Type Ia supernova data and between $H_0$ and the $r_d$ for the BAO data with a new degeneracy between $r_d$ and $M_B$. By choosing two different values for $r_d^{}$, one specifies the $M_B$ from NRT. The sound horizon from Planck $r_d^\mathrm{Pl18} = 147.09\pm0.26\,\mathrm{Mpc}$ \cite{Planck:2018vyg} gives $M_B = -19.38\pm0.20$ from ANN and $M_B = -19.42\pm0.35$ from GPs. The sound horizon from H0LiCOW+SN+BAO+SH0ES $r_d^\mathrm{HW+SN+BAO+SH0ES} = 136.1\pm2.7\,\mathrm{Mpc}$ \cite{Arendse:2019hev} $M_B=-19.22\pm0.20$ from ANN and  $M_B = -19.25\pm0.39$ from GPs. The latter value of $M_B$ matches the recently measured value by Ref.~\cite{Riess:2021jrx}. These mean values have been obtained assuming one Gaussian distribution for the whole reconstruction. For the ANN, however, one can fit the distribution with a bimodal one (see Table \ref{ANNGP}), which suggests two sub-sets of preferred values for $M(z)$. The GP can be fit with a Gaussian with a tail, which when accounted for makes the mean values a bit lower than the cited above. 

From the NRT approaches, we see a $1 \sigma$ possibility of a nuisance parameter presence ($\delta M_B(z)\neq0$) in higher redshifts. The two different approaches (ANN and GP) differ in their error predictions, with the ANN a lot less noisier, but both of them seem to predict some kind of deviation from $M_B=const$ at $z>1$. One possible reason for this deviation could be that it is due to the big errors of the high-redshift BAO points. For this reason, on our plots, we have removed the last two BAO points, so that we can avoid the large errors of the reconstruction.
In our results we see that the two NRT approaches have different behavior of $M_B$ for small redshifts ($z\in (0,0.25)$). This is due to the initial numerical singularity that GPs has, but the ANN does not. 
Over-all, the two different methods show very similar behaviour in the range $z\in [0.25,2]$ -- fluctuations and a decreasing trend with the redshift with an increasing error due to the low number of datapoints for higher redshifts.

Using Bayesian Analysis we test different phenomenological models for $\delta M(z)$ and we see that there is no strong preference for any model, despite the constant model of $M_B$ giving the strongest constraints on the parameters. This means that the question of whether there is a better functional dependence describing $M_B(z)$ is open and that we need better high-redshift data to be able to get strong statistical preferences for any model. The conclusion from our application of different approaches on the combined SN and BAO datasets is that the constancy of $M_B$ we observe is within $1\sigma$ which leaves the door open for a possibility for a nontrivial nuisance parameter presence $\delta M_B(z) <0$ in higher redshifts.

\acknowledgments 
We thank Eleonora Di-Valentino, Sunny Vagnozzi and Blake Sherwin for useful comments and discussions. D.B gratefully acknowledge the support the supports of the Blavatnik and the Rothschild fellowships. D.B. acknowledges a Postdoctoral Research Associateship at the Queens' College, University of Cambridge. We have received partial support from European COST actions CA15117 and CA18108 and STFC consolidated grants ST/P0006811 and ST/T0006941. D.S. is thankful to Bulgarian National Science Fund for support via research grants KP-06-N58/5. The authors would like to acknowledge networking support by the COST Action CA18108 and funding support from IntelliVerse which is supported by the Malta Digital Innovation Authority. This research has been carried out using computational facilities procured through the European Regional Development Fund, Project No. ERDF-080 "A supercomputing laboratory for the University of Malta".

\bibliographystyle{apsrev4-1}
\bibliography{ref}

\begin{appendix}
\section{Details on the MCMC}

\clearpage
\begin{table*}[h!]
	\begin{center}
		\begin{tabular}{|c|c|c|c|c|c|c|c|c|}
			\hline
               \multicolumn{9}{|c|}{$r^{Pl18}_d = 147.09 \pm 0.26$ Mpc}\\  
            \hline
			Model & $H_0$ & $\Omega_m$ & $\Omega_r$ & $M$ & $\epsilon$ & $\delta$ & $\Omega_\Lambda$ & $r_d$ \\
			\hline
			$\epsilon=0$ & $69.6\pm 0.49$ & $0.3\pm 0.01$ & $0.006\pm 0.003$ & $-19.21\pm 0.03$ & 0.000 & 0 & $0.69\pm 0.01$ & $147.1\pm 0.3$ \\
			\hline
			Line & $68.71\pm 0.73$ & $0.33\pm 0.02$ & $0.0054\pm 0.0033$ & $-19.21\pm 0.03$ & $0.06\pm 0.03$ & $0.17\pm 0.01$ & $0.66\pm 0.02$ & $147.1\pm 0.3$ \\
			\hline
			A & $68.7\pm 0.67$ & $0.34\pm 0.02$ & $0.0058\pm 0.0036$ & $-19.22\pm 0.04$ & $0.35\pm 0.24$ & $0.32\pm 0.2$ & $0.66\pm 0.02$ & $147.1\pm 0.2$ \\
			\hline
			B & $68.71\pm 0.68$ & $0.34\pm 0.03$ & $0.0058\pm 0.0033$ & $-19.23\pm 0.03$ & $0.13\pm 0.06$ & $0.29\pm 0.22$ & $0.66\pm 0.02$ & $147.1\pm 0.3$ \\
			\hline
			C & $68.59\pm 0.78$ & $0.34\pm 0.03$ & $0.0058\pm 0.0035$ & $-19.23\pm 0.03$ & $0.11\pm 0.05$ & $0.52\pm 0.32$ & $0.65\pm 0.02$ & $147.1\pm 0.3$ \\
			\hline
            \multicolumn{9}{|c|}{$r_d^{HW+SN+BAO+SH0ES} =136.1 \pm  2.7$ Mpc}\\
            \hline
			$\epsilon=0$ & $72.65\pm 1.09$ & $0.31\pm 0.01$ & $0.0064\pm 0.0025$ & $-19.2\pm 0.03$ & 0.000 & 0 & $0.69\pm 0.01$ & $140.4\pm 1.8$ \\
			\hline
			Line & $71.17\pm 1.34$ & $0.35\pm 0.02$ & $0.0052\pm 0.0033$ & $-19.2\pm 0.02$ & $0.09\pm 0.03$ & $0.16\pm 0.01$ & $0.64\pm 0.02$ & $140.8\pm 2.3$ \\
			\hline
			A & $72.2\pm 1.24$ & $0.34\pm 0.02$ & $0.0062\pm 0.003$ & $-19.2\pm 0.03$ & $0.29\pm 0.19$ & $0.43\pm 0.27$ & $0.65\pm 0.02$ & $139.5\pm 2.4$ \\
			\hline
			B & $73.63\pm 1.13$ & $0.33\pm 0.02$ & $0.0061\pm 0.0032$ & $-19.23\pm 0.03$ & $0.26\pm 0.08$ & $0.07\pm 0.05$ & $0.67\pm 0.02$ & $137.7\pm 2.6$ \\
			\hline
			C & $74.1\pm 1.64$ & $0.33\pm 0.02$ & $0.0059\pm 0.0026$ & $-19.24\pm 0.03$ & $0.27\pm 0.11$ & $0.09\pm 0.07$ & $0.66\pm 0.03$ & $136.7\pm 2.6$ \\
            \hline
            \multicolumn{9}{|c|}{$\Omega_b = 0.0224 \pm 0.0001$}\\
            \hline
			$\epsilon=0$ & $64.74\pm 2.66$ & $0.3\pm 0.01$ & $0.004\pm 0.0003$ & $-19.24\pm 0.03$ & 0.000 & 0 & $0.69\pm 0.01$ & $158.6\pm 6.6$* \\
			\hline
			Line & $65.48\pm 2.68$ & $0.34\pm 0.03$ & $0.004\pm 0.0003$ & $-19.24\pm 0.04$ & $0.06\pm 0.04$ & $0.17\pm 0.01$ & $0.66\pm 0.03$ & $154.9\pm 6.3$* \\
			\hline
			A & $64.97\pm 2.35$ & $0.33\pm 0.03$ & $0.004\pm 0.0003$ & $-19.24\pm 0.03$ & $0.33\pm 0.29$ & $0.29\pm 0.21$ & $0.67\pm 0.03$ & $156.3\pm 5.3*$ \\
			\hline
			B & $66.56\pm 3.6$ & $0.33\pm 0.02$ & $0.0038\pm 0.0004$ & $-19.25\pm 0.04$ & $0.11\pm 0.08$ & $0.34\pm 0.28$ & $0.66\pm 0.02$ & $152.8\pm 7.4$ \\
			\hline
			C & $68.34\pm 4.26$ & $0.33\pm 0.03$ & $0.0037\pm 0.0008$ & $-19.24\pm 0.03$ & $0.14\pm 0.11$ & $0.32\pm 0.3$ & $0.66\pm 0.03$ & $149.8\pm 14.9*$ \\
             \hline
		\end{tabular}
	\end{center}
	\caption{The final values of the inferred parameters for the considered models of nuisance parameter and with the different ways to set $r_d$ discussed in the text. Values marked with $*$ are calculated, not inferred}
	\label{cosmo1}
\end{table*}

\begin{table}[!htb]
	\begin{center}
		\begin{tabular}{|c|c|c|c|c|c|c|c|}
			\hline
            \multicolumn{8}{|c|}{$r^{Pl18}_d = 147.09 \pm 0.26$ Mpc} \\
            \hline
			Model & AIC & $\Delta$AIC & BIC & $\Delta BIC$ & DIC & $\Delta$DIC & log(BF) \\
			\hline
			$\epsilon=0$ & 84.9 &  & 99.7 &  & 63.7 &  &  \\
			\hline
			Line & 87.5 & -2.6 & 103.4 & -3.7 & 63.4 & 0.29 & 1.5 \\
			\hline
			A & 90.2 & -5.3 & 107.1 & -7.4 & 63.1 & 0.65 & 0.4 \\
			\hline
			B & 90.1 & -5.2 & 107.0 & -7.4 & 62.9 & 0.72 & 0.9 \\
			\hline
			C & 89.9 & -5.1 & 106.9 & -7.2 & 62.9 & 0.85 & 0.4 \\
			\hline
            \multicolumn{8}{|c|}{$r_d^{HW+SN+BAO+SH0ES} =136.1 \pm  2.7$ Mpc}\\
			\hline
			$\epsilon=0$ & 85.2 &  & 100.0 &  & 64.1 &  &  \\
			\hline
			Line & 87.9 & -2.8 & 103.9 & -3.9 & 64.0 & 0.04 & 1.3 \\
			\hline
			A & 90.3 & -5.1 & 107.2 & -7.2 & 63.2 & 0.9 & -1.4 \\
			\hline
			B & 89.8 & -4.6 & 106.7 & -6.7 & 62.7 & 1.4 & -3.4 \\
			\hline
			C & 90.4 & -5.2 & 107.3 & -7.3 & 63.3 & 0.8 & -2.3 \\
            \hline
            \multicolumn{8}{|c|}{$\Omega_b = 0.0224 \pm 0.0001$}\\
            \hline
            $\epsilon=0$ & 85.1 &  & 99.9 &  & 63.9 &  &  \\
			\hline
			Line & 87.8 & -2.7 & 103.7 & -3.8 & 63.8 & 0.1 & 2.9 \\
			\hline
			A & 90.6 & -5.5 & 107.5 & -7.6 & 63.5 & 0.4 & 1.0 \\
			\hline
			B & 90.3 & -5.1 & 107.1 & -7.2 & 63.1 & 0.8 & 1.6 \\
			\hline
			C & 90.4 & -5.3 & 107.4 & -7.5 & 63.4 & 0.6 & 1.6 \\
			\hline
		\end{tabular}
	\end{center}
	\caption{\label{tab:mcmc_vals2_rd2} \it{The selection criteria values for different models of nuisance parameter }}
 \label{selcrit}
\end{table}

\end{appendix}
\end{document}